\documentclass[epj]{svjour}

\usepackage[english]{babel}
\usepackage{graphicx}
\graphicspath{{./}}

\newcommand{\vep}{\varepsilon}
\newcommand{\als}{\alpha_s}
\newcommand{\nf}{n_f}
\newcommand{\nl}{n_l}
\newcommand{\nh}{n_h}
\newcommand{\Cb}{\overline{C}}
\newcommand{\Log}[2]{\log^{#2}({#1})}
\newcommand{\z}[1]{\zeta_{#1}}
\newcommand{\Order}{\mathcal{O}}

\renewcommand{\*}{\,}

\newcommand{\ice}[1]{\relax}
\newcommand{\GeV}{\, \mbox{GeV}}

\begin{document}
\title{Four-loop moments of the heavy quark vacuum polarization 
          function in perturbative QCD}

\author{K.~G.~Chetyrkin\inst{1} 
\thanks{On leave from Institute for Nuclear Research of the 
        Russian Academy of Sciences, Moscow, 117312, Russia.}%
        \and 
        J.~H.~K\"uhn\inst{1} 
        \and C.~Sturm\inst{2}
}                     
\institute{
Institut f{\"u}r Theoretische Teilchenphysik,
Universit{\"a}t Karlsruhe,
D-76128 Karlsruhe, 
Germany
\and 
Dipartimento di Fisica Teorica, 
Universit{\`a} di Torino, I-10125 Torino, Italy 
{\rm{\&}}
INFN, Sezione di Torino, Italy}
\date{\mbox{}}
\abstract{ New results at four-loop order in perturbative QCD for the
  first two Taylor coefficients of the heavy quark vacuum polarization
  function are presented.  They can be used to perform a precise
  determination of the charm- and bottom-quark mass. Implications for
  the value of the quark masses are briefly discussed.
\PACS{
      {12.38.Bx}{Perturbative calculations}\and
      {12.38.-t}{Quantum chromodynamics}\and
      {11.55.Fv}{Dispersion relations}\and
      {14.65.Dw}{Charmed quarks}\and
      {14.65.Fy}{Bottom quarks}
     } 
} 
\maketitle
\section{Introduction}
\label{Intro}
Two-point correlators are of central importance for many theoretical and
phenomenological investigations in Quantum Field Theory. As a
consequence they are studied in great detail in the framework of
perturbative calculations. Due to simple kinematics (only one external
momentum) even multi-loop calculations can be performed. The results for
all physically interesting diagonal and non-diagonal correlators
including {\em full} quark mass dependence are available up to
$\mathcal{O}\left(\alpha_s^2\right)$
\cite{Chetyrkin:1996cf,Chetyrkin:1998ix,Chetyrkin:1997mb}.\\
At four-loop order the two-point correlators can be considered in two
limits: In the high energy limit massless propagators need to be
calculated, in the low energy limit vacuum integrals (``tadpole
diagrams'') arise. The evaluation of the latter in three-loop
approximation has been pioneered in ref. \cite{Broadhurst:1992fi} and
automated in ref. \cite{Steinhauser:2000ry}.\\
Recently first results for physical quantities, which are related to
four-loop tadpole diagrams, have been obtained.  The four-loop matching
condition for the strong coupling constant $\alpha_s$ at a heavy quark
threshold has been calculated in
ref.~\cite{Chetyrkin:2005ia,Schroder:2005hy}.  The four-loop QCD
contribution to the electroweak $\rho$-parameter induced by the singlet
diagrams of the Z-boson self-energy has been computed in
ref.~\cite{Schroder:2005db}.\\
The detailed knowledge of the heavy quark correlator is important for
the precise determination of heavy quark masses with the help of QCD sum
rules. As known from ref.~\cite{Kuhn:2001dm,Kuhn:2002zr}, the
determination of the charm- and bottom-quark mass is further improved,
if the four-loop corrections, hence $\Order(\als^3)$, for the lowest
Taylor coefficients of the vacuum polarization function are
available. The subset of four-loop contributions to the lowest two
moments, which involve two internal loops from massive and massless
fermions coupled to gluons, hence of $\Order(\als^3\*\nf^2)$, has
already been calculated in ref.~\cite{Chetyrkin:2004fq}. The symbol
$\nf$ denotes the number of active quark-flavors, contributing through
fermion-loops inserted into gluon lines.  The terms being proportional
to $\als^j\*\nl^{j-1}$ are even known to all orders $j$ in perturbative
QCD~\cite{Grozin:2004ez}. The symbol $\nl$ denotes the number of light
quarks, considered as massless. In this paper the complete four-loop
contributions originating from non-singlet diagrams for the first two
Taylor coefficients are presented.  Singlet contributions have been
studied in ref.~\cite{Groote:2001vr,Portoles:2001yu,Portoles:2002rt}.\\
In general, the tadpole diagrams encountered in these calculations
contain both massive and massless lines. As is well-known, the
computation of the four-loop $\beta$-function can be reduced to the
evaluation of four-loop tadpoles composed of { completely massive}
propagators only.  Calculations for this case have been performed in
\cite{vanRitbergen:1997va,Czakon:2004bu,Kajantie:2003ax}.

The outline of this paper is as follows. In section
\ref{GeneralNotations} we briefly introduce the notation and discuss
generalities. In section \ref{Calculations} we discuss the reduction to
master integrals, describe the solution of the linear system of
equations, give the result for the lowest two moments and discuss briefly  the
impact on the quark mass determination.  Our conclusions and a brief
summary are given in section \ref{DiscussConclude}.

\section{Notation and Generalities}
\label{GeneralNotations}
The correlator $\Pi^{\mu\nu}(q)$ of two currents is defined as
\begin{equation}
  \Pi^{\mu\nu}(q,j)=i\*\int dx\,e^{iqx}\langle 0|Tj^\mu(x) j^\nu(0)|0
  \rangle\,, 
\label{correl}
\end{equation}
with the current $j^{\mu}(x)=\overline{\Psi}(x)\gamma^\mu\Psi(x)$ being
composed out of the heavy quark fields $\Psi(x)$. The function
$\Pi^{\mu\nu}(q)$ is conveniently written in the form:
\begin{equation}
  \label{vacpol}
  \Pi^{\mu\nu}(q)=\left(-q^2\*g^{\mu\nu}+q^{\mu}\*q^{\nu}\right)\*\Pi(q^2)\,.
\end{equation}
The vanishing of the longitudinal contribution, as well as the
confirmation of $\Pi^{\mu}_{\phantom{\mu}\mu}(q^2=0)=0$, has been used
as a check of the calculation. The function $\Pi(q^2)$ is of
phenomenological interest, because it can be related to the ratio
$R(s)=\sigma(e^+\*e^-\rightarrow\mbox{hadrons})/\sigma(e^+\*e^-\rightarrow\mu^+\mu^-)$
with the help of dispersion relations:
\begin{equation}
\Pi(q^2)={1\over12\*\pi^2}\int_{0}^{\infty}\!ds\,{R(s)\over (s-q^2)}\;\;\mbox{mod subtr.}
\label{PIR}
\end{equation}
Performing the derivative of eq.~(\ref{PIR}) with respect to $q^2$ one
obtains on the one hand ``experimental'' moments
\begin{equation}
\mathcal{M}^{\mbox{\footnotesize{exp}}}_n=\int\!ds\,{R(s)\over s^{n+1}}\,,
\label{Mexp}
\end{equation}
which can be evaluated from the $R$-ratio. On the other hand one can
define ``theoretical'' moments
\begin{equation}
\mathcal{M}^{\mbox{\footnotesize{th}}}_n=
Q_{q}^2\*{9\over4}\*\left({1\over4\*\overline{m}_q^2}\right)^n\*\overline{C}_n\,,
\label{Mth}
\end{equation}
which are related to the Taylor coefficients $\overline{C}_n$ of the
vacuum polarization function:
\begin{equation}
  \overline{\Pi}(q^2) = {3\*Q_q^2\over16\*\pi^2}\*\sum_{n\ge0} 
                        \overline{C}_n\*\overline{z}^n\,,
\label{PiExp}
\end{equation}
with $\overline{z}=q^2/(4\*\overline{m}^2)$. Symbols carrying a bar
denote that renormalization has been performed in
$\overline{\mbox{MS}}$-scheme. The Taylor expansion in $q^2$ around
$q^2=0$ leads to massive tadpole integrals.  The first and higher
derivatives can be used for a precise determination of the charm- and
bottom-quark mass. However, also the lowest expansion coefficient
$\overline{C}_0$ has an interesting physical meaning, since it relates
the coupling of the electromagnetic interaction in different
renormalization schemes. \\
Sum rules as tool for the determination of the charm- and bottom-quark
mass have been suggested since long in ref.~\cite{Novikov:1977dq}. This
method has then been applied later also to the determination of the
bottom-quark mass~\cite{Reinders:1984sr}. One of the most precise
determinations of the charm- and bottom-quark mass being based on sum
rules in connection with the calculation of three-loop moments in
perturbative QCD has been performed in ref.~\cite{Kuhn:2001dm}, with the
values of $\overline{m}_c(\overline{m}_c)=1.304(27)$~GeV and
$\overline{m}_b(\overline{m}_b)=4.191(51)$~GeV as results for the charm-
and the bottom-quark mass.

It is convenient to define the expansion of the Taylor coefficients
$\overline{C}_n$ of the vacuum polarization function in the strong
coupling constant $\als$ as
\begin{equation}
\Cb_n=\Cb_n^{(0)}
   +\left({\als\over\pi}\right)^1\*\Cb_n^{(1)}
   +\left({\als\over\pi}\right)^2\*\Cb_n^{(2)}
   +\left({\als\over\pi}\right)^3\*\Cb_n^{(3)}+\dots\,.
\label{CbarExp}
\end{equation}
Due to the distinct mass hierarchy of the quarks in the Standard-Model,
one can consider for the above moments one species of quarks as massive
($\nh=1$) and all lighter ones as massless. The number of active quarks
$\nf$ is then decomposed according to $\nf=\nl+\nh$.  For convenience
the symbol $\nh$ is kept explicitly in the following.
\section{Calculations and Results}
\label{Calculations}
\begin{figure}[!ht]
\begin{center}
\begin{minipage}[b]{2.5cm}
  \begin{center}
    \includegraphics[height=2cm,bb=126 332 460 665]{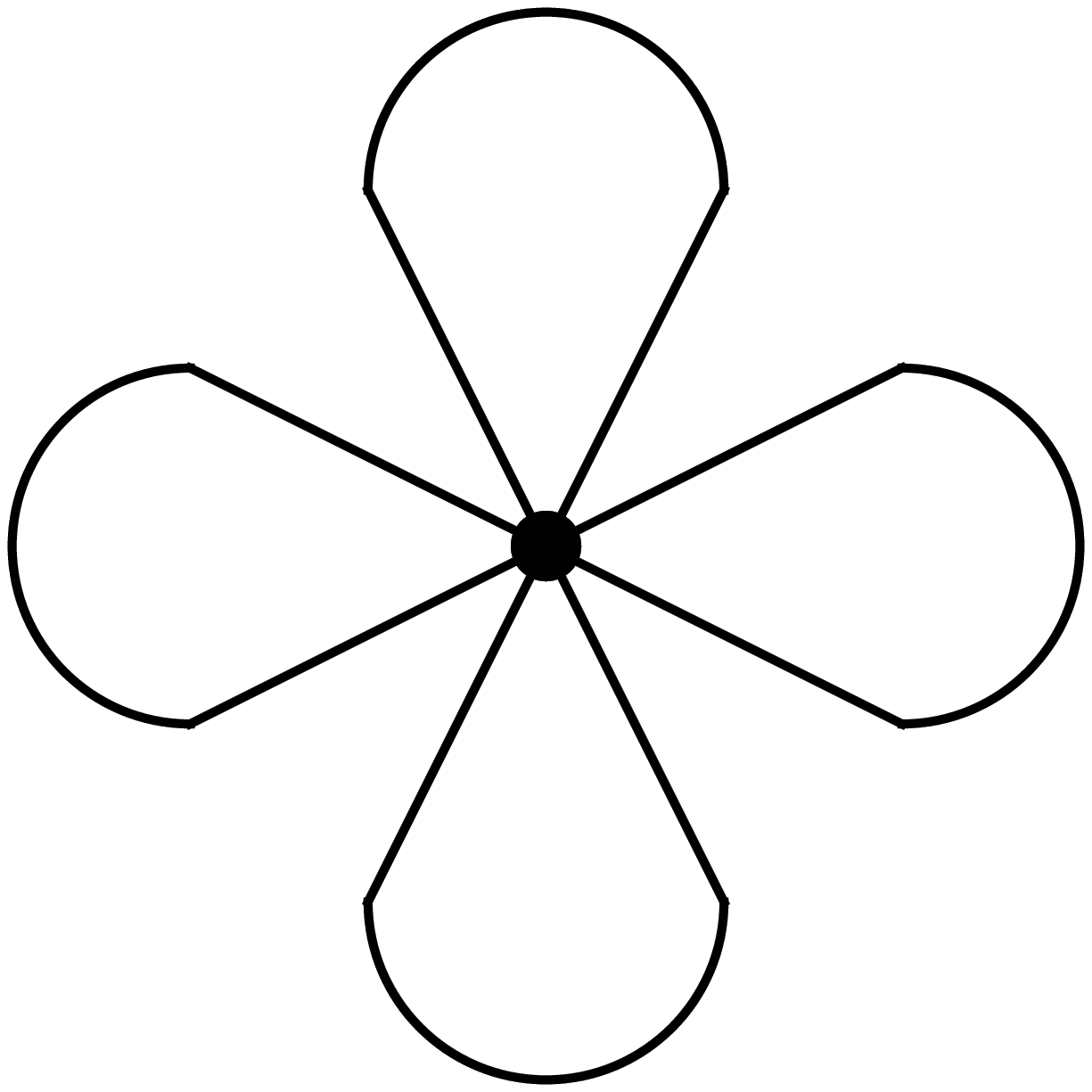}\\[0.5ex]
\hspace*{0.4cm}$\mathrm{T}_{41}$
  \end{center}
\end{minipage}
%
%
\begin{minipage}[b]{2.5cm}
  \begin{center}
    \includegraphics[height=2cm,bb=170 320 415 666]{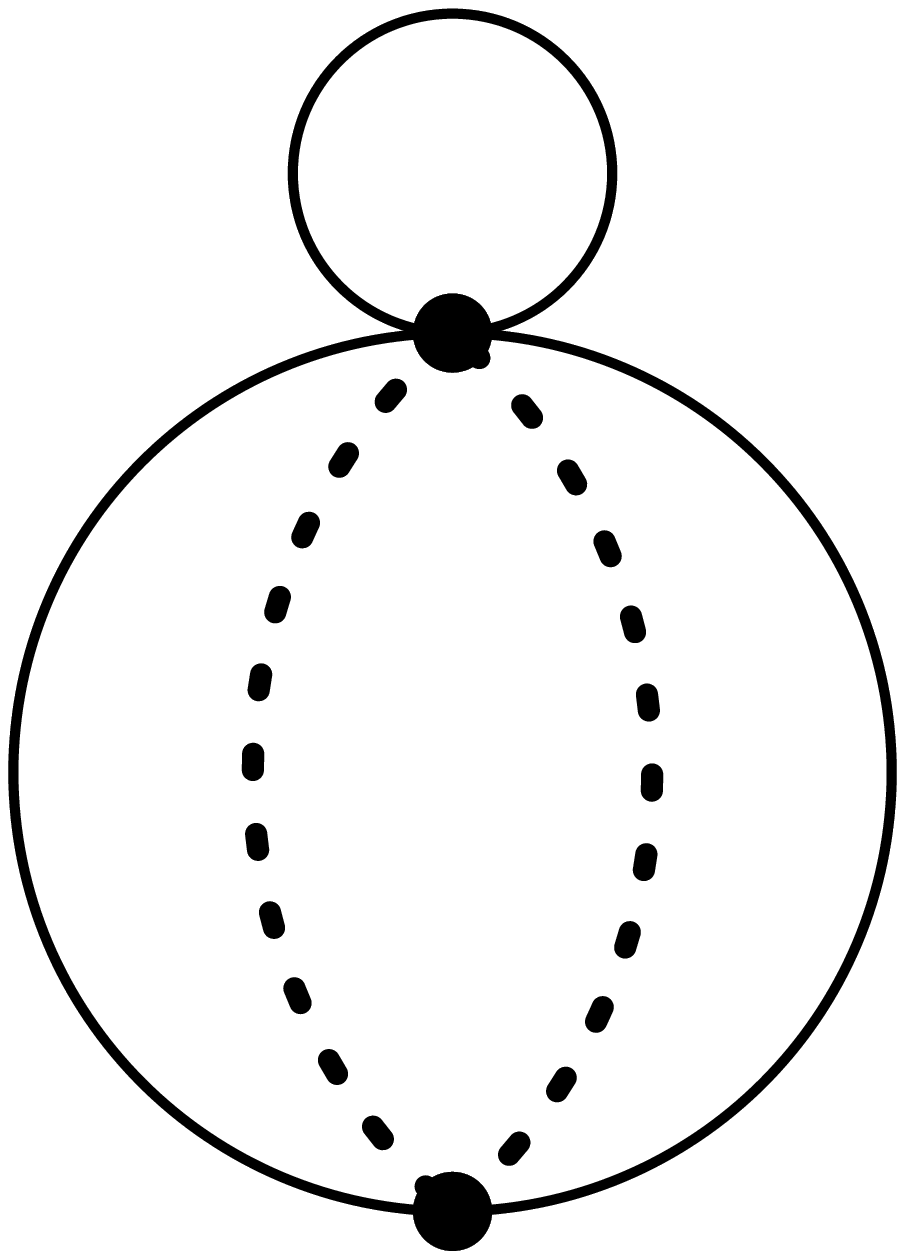}\\[0.5ex]
\hspace*{0.4cm}$\mathrm{T}_{51}$
  \end{center}
\end{minipage}
%
%
\begin{minipage}[b]{2.5cm}
  \begin{center}
    \includegraphics[height=2cm,bb=126 320 460 678]{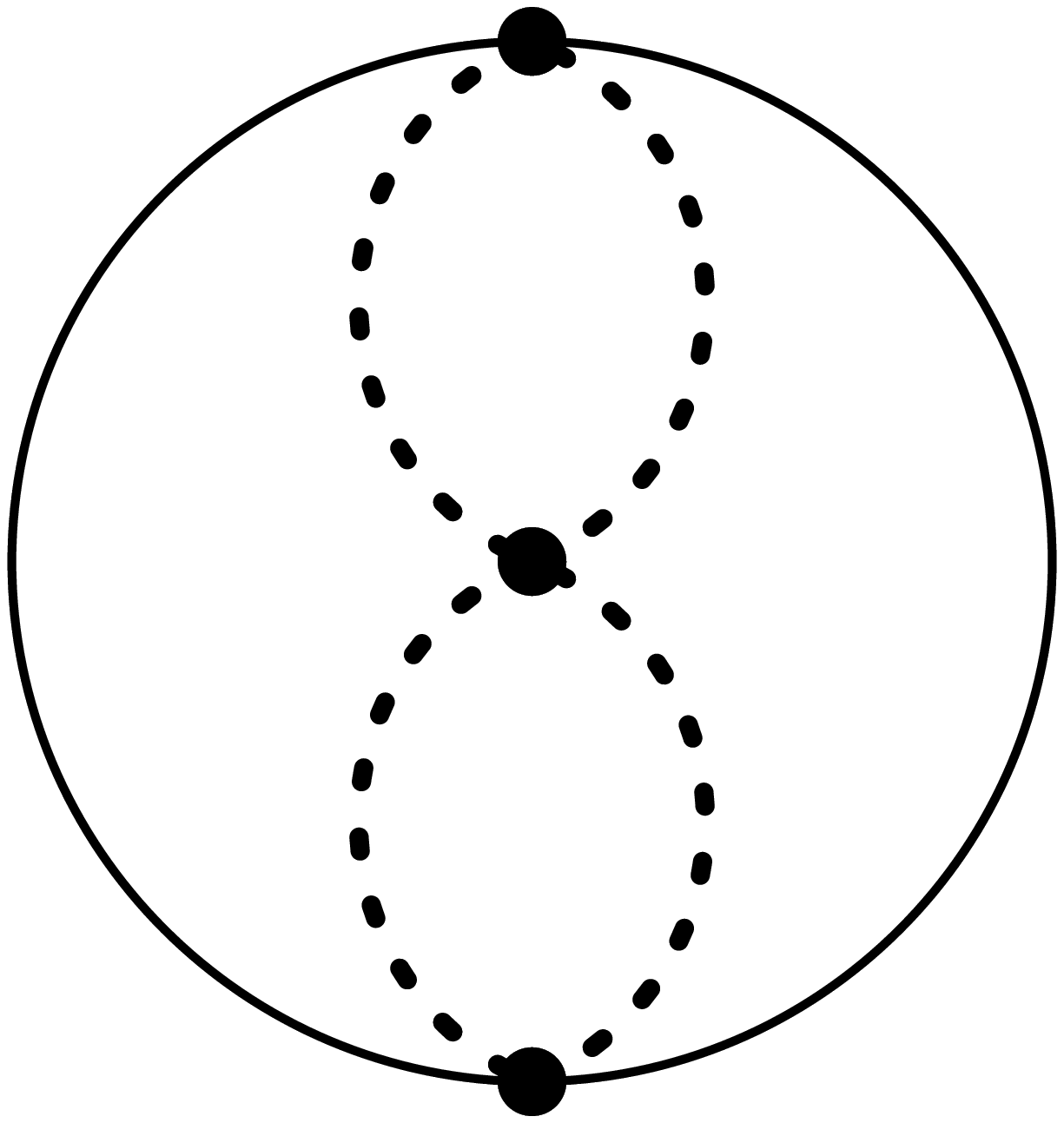}\\[0.5ex]
\hspace*{0.4cm}$\mathrm{T}_{63}$
  \end{center}
\end{minipage}
%
%
\\[2ex]
%
\begin{minipage}[b]{2.5cm}
  \begin{center}
    \includegraphics[height=2cm,bb=126 320 460 678]{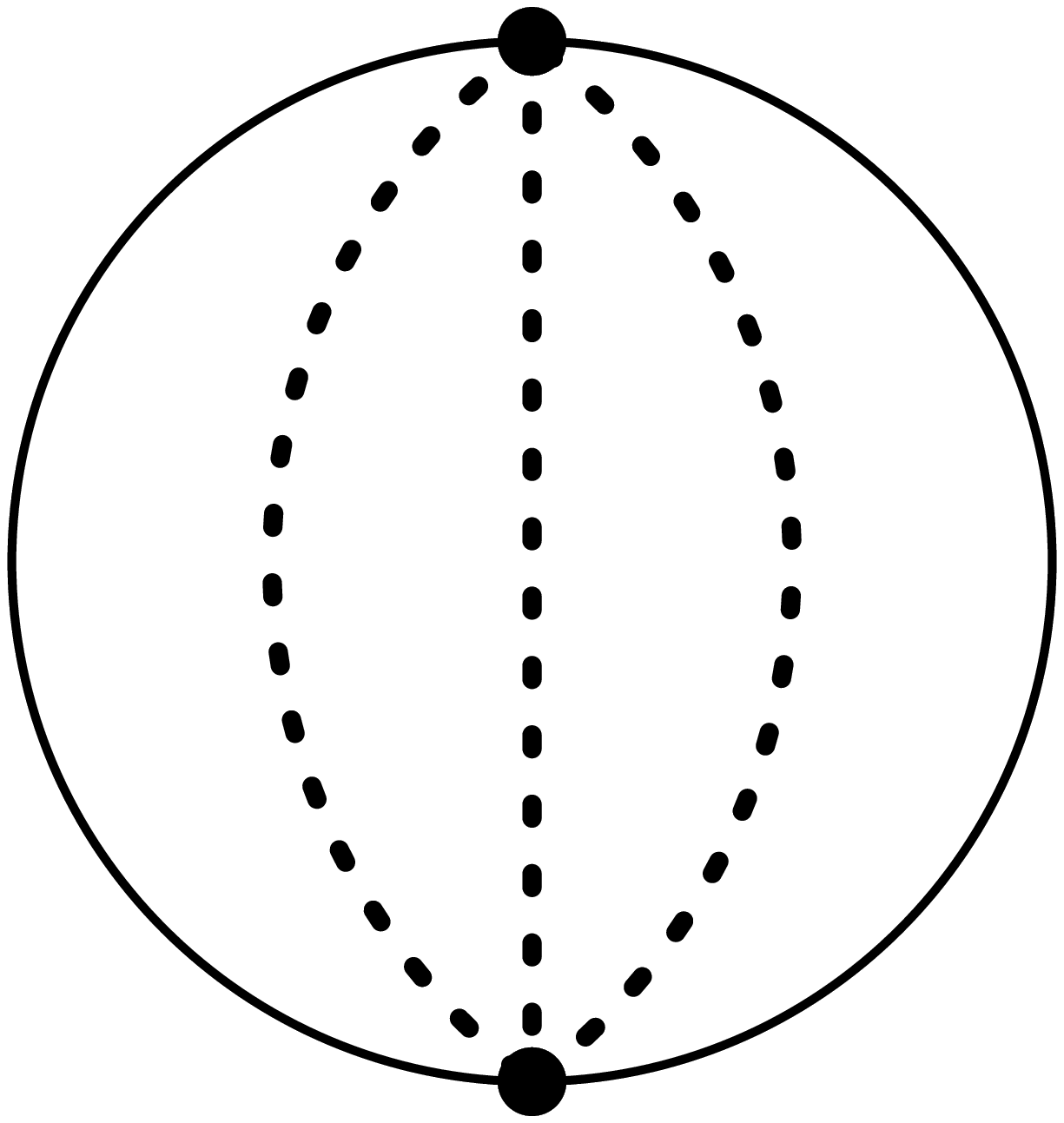}\\[0.5ex]
\hspace*{0.4cm}$\mathrm{T}_{53}$
  \end{center}
\end{minipage}
\begin{minipage}[b]{2.5cm}
  \begin{center}
    \includegraphics[height=2cm,bb=170 320 415 666]{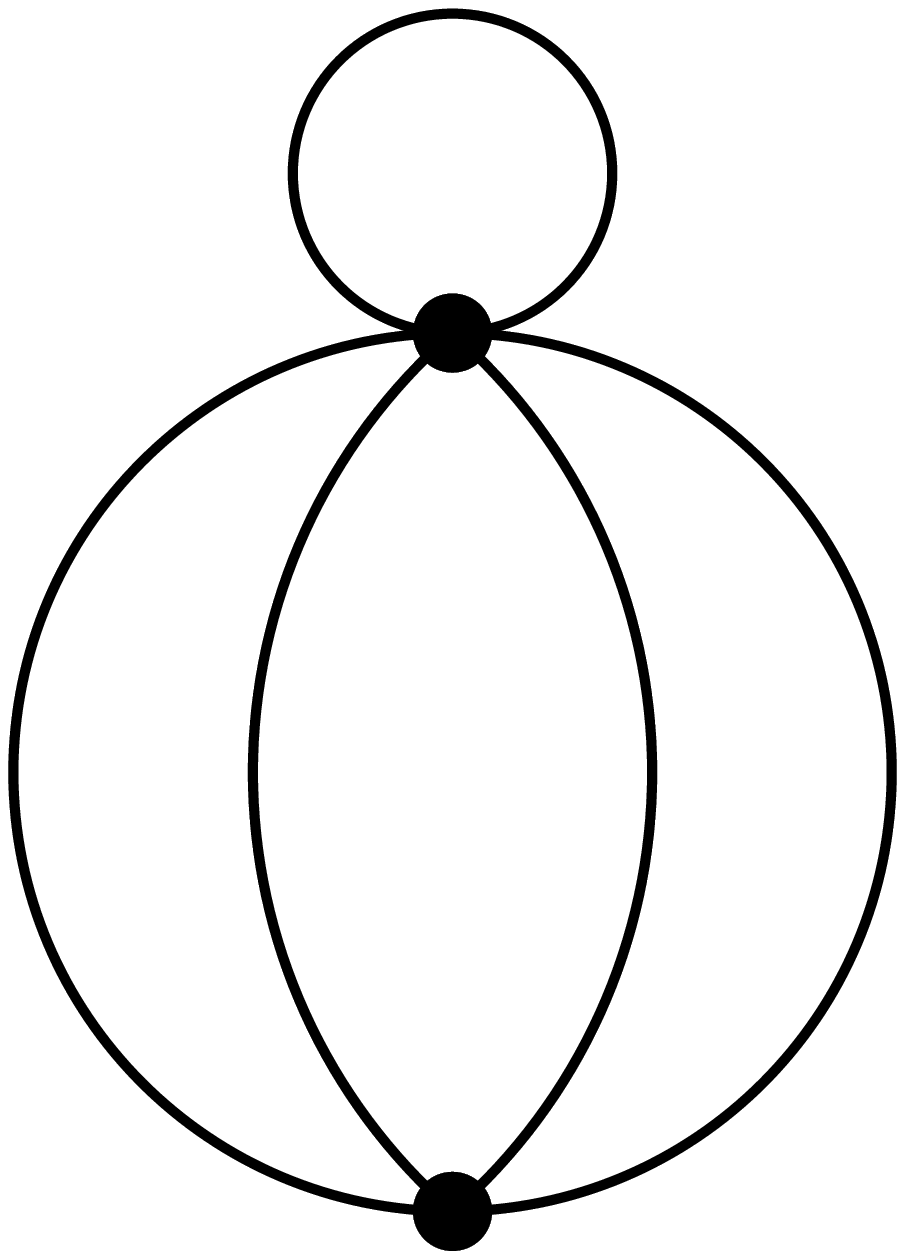}\\[0.5ex]
\hspace*{0.4cm}$\mathrm{T}_{52}$
  \end{center}
\end{minipage}
%
%
%
 \end{center}
\caption{Factorized or analytically known master integrals. The solid
(dashed) lines denote massive (massless) propagators.
\label{fig:1}} 
\end{figure}

The Feynman-diagrams have been generated with the help of the program
{\tt{QGRAF}} \cite{Nogueira:1991ex}. After performing the expansion in
the external momentum $q$ all integrals can be expressed in terms of 55
independent vacuum topologies with additional increased powers of
propagators and additional irreducible scalar products.\\
In order to reduce this host of integrals to a small set of master
integrals the traditional Integration-by-parts(IBP) method has been used
in combination with Laporta's algorithm
\cite{Laporta:1996mq,Laporta:2001dd}.  The resulting system of linear
equations has been solved with a {\tt{FORM3}}
\cite{Vermaseren:2000nd,Vermaseren:2002rp,Tentyukov:2006ys} based
program, in which partially also ideas described in
ref.~\cite{Laporta:2001dd,Mastrolia:2000va,Schroder:2002re} have been
implemented. The rational functions in the space-time dimension $d$,
which arise in this procedure, have been simplified with the program
{\tt{FERMAT}}~\cite{Lewis}. Masking of large integral coefficients has
been implemented, a strategy also adopted in the program
{\tt{AIR}}~\cite{Anastasiou:2004vj}. In order to achieve the reduction
to 13 master integrals more than 31 million IBP-equations have been
generated and solved. This leads to integral-tables with solutions for
around five million integrals.  Furthermore all symmetries of the 55
independent topologies have been taken into account in an automated way,
by reshuffling the powers of the propagators of a given topology in a
unique way. Taking into account symmetries is important in order to keep
the size of the integral-tables under control.\\
The first four master integrals shown in fig.~\ref{fig:1} can be
calculated completely analytically in terms of $\Gamma$-functions.  The
fifth integral ($T_{52}$) can be obtained from results of
ref.~\cite{Broadhurst:1992fi,Broadhurst:1996az}.

\begin{figure}[!ht]
 \begin{center}
\begin{minipage}[b]{2.5cm}
  \begin{center}
    \includegraphics[height=2cm,bb=126 320 460 678]{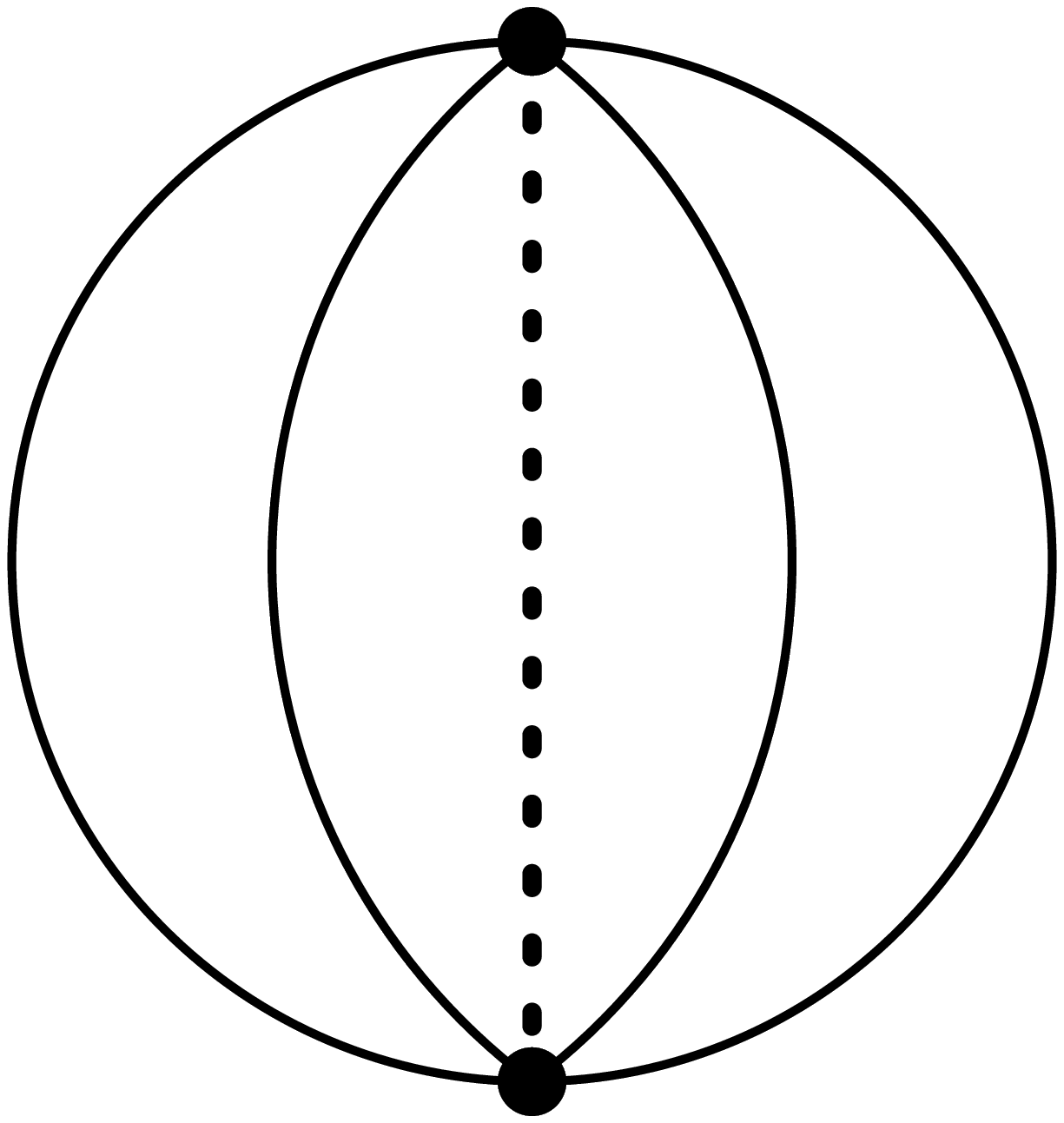}\\[0.5ex]
\hspace*{0.4cm}$\mathrm{T}_{54}$
  \end{center}
\end{minipage}
%
\begin{minipage}[b]{2.5cm}
  \begin{center}
    \includegraphics[height=2cm,bb=126 320 460 678]{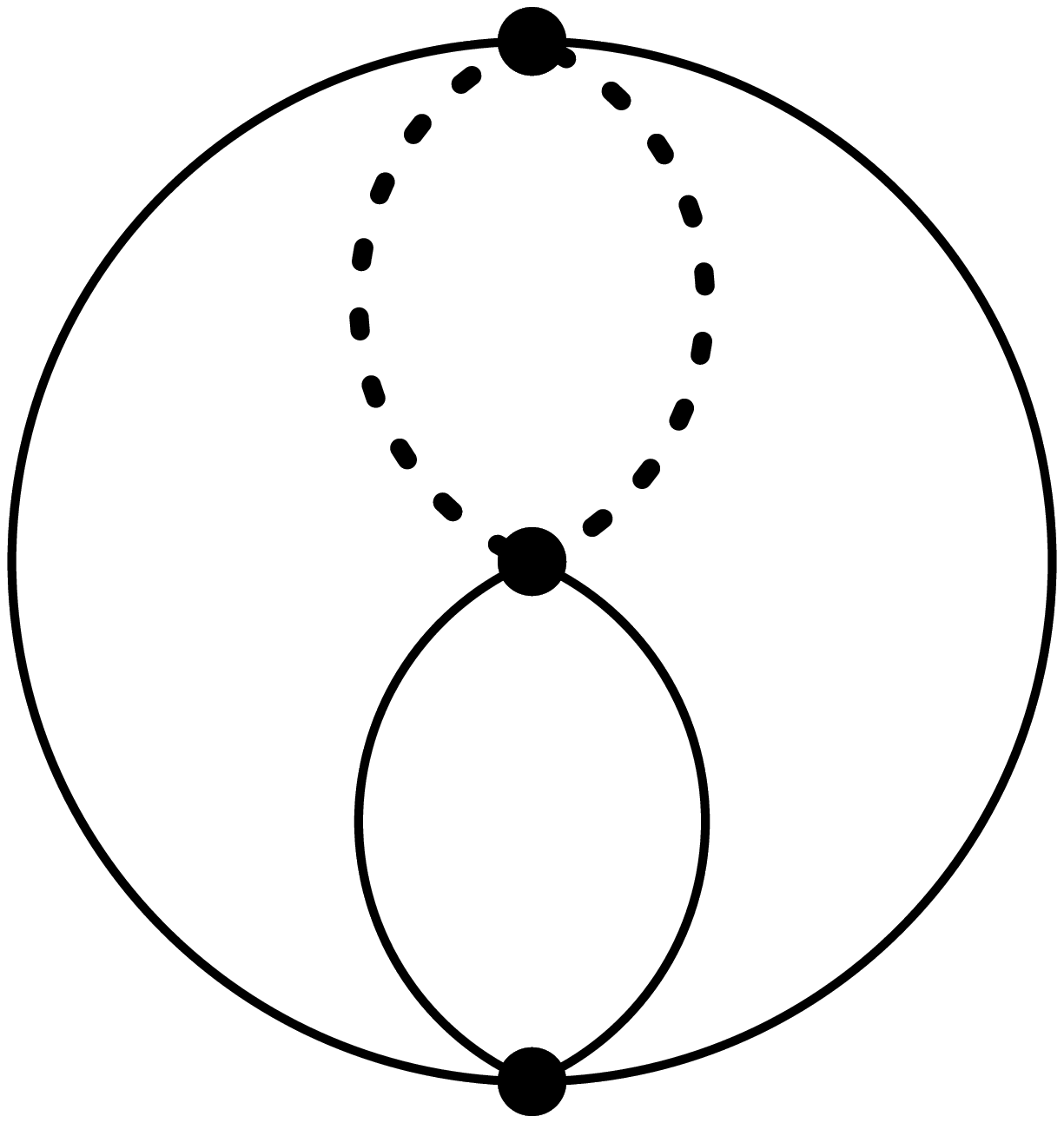}\\[0.5ex]
\hspace*{0.4cm}$\mathrm{T}_{62}$
  \end{center}
\end{minipage}
%
%
\begin{minipage}[b]{2.5cm}
  \begin{center}
    \includegraphics[height=2cm,bb=126 320 460 678]{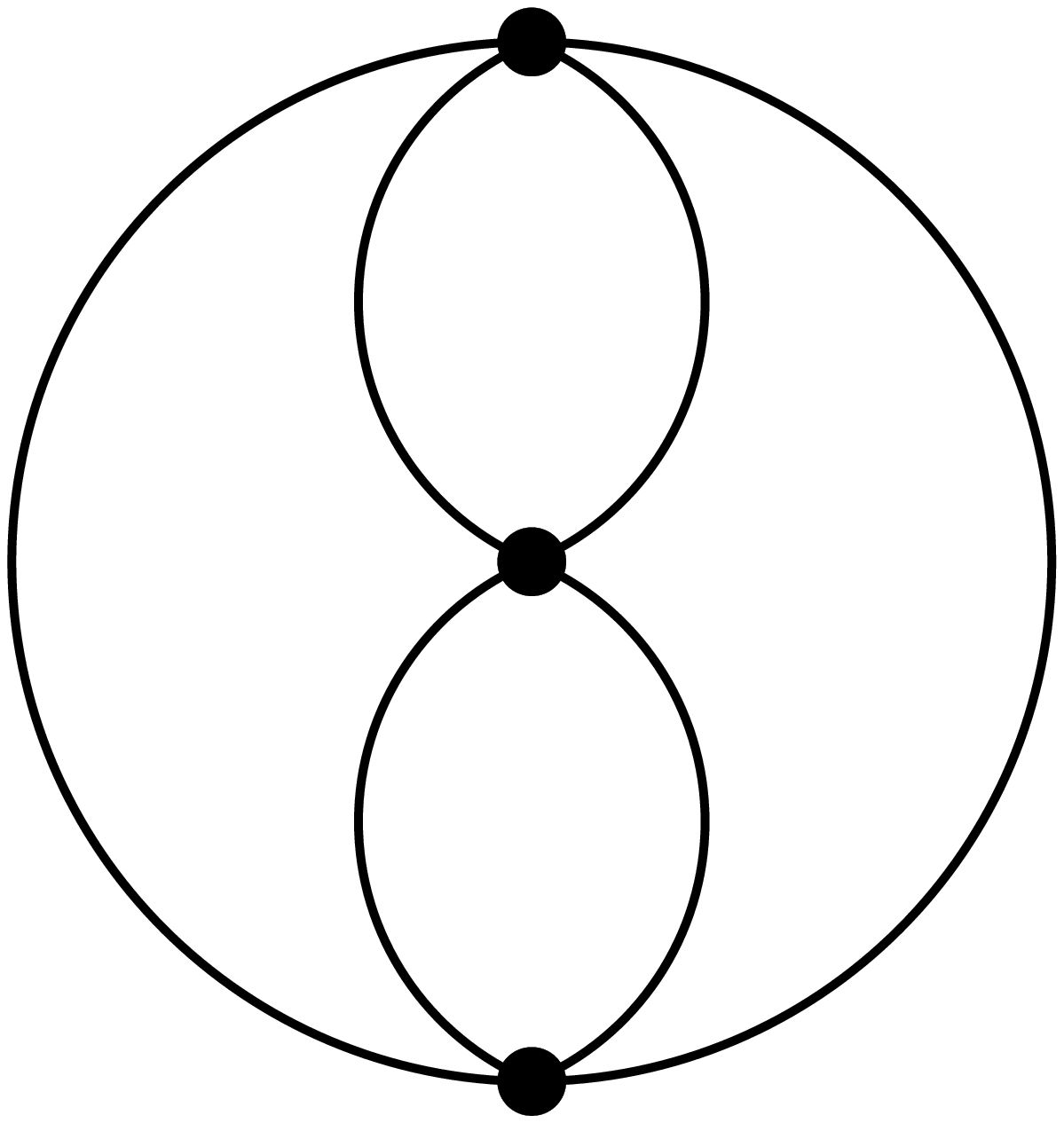}\\[0.5ex]
\hspace*{0.4cm}$\mathrm{T}_{61}$
  \end{center}
\end{minipage}
%
%
\\[2ex]
%
%
\begin{minipage}[b]{2.5cm}
  \begin{center}
    \includegraphics[height=2cm,bb=126 332 460 678]{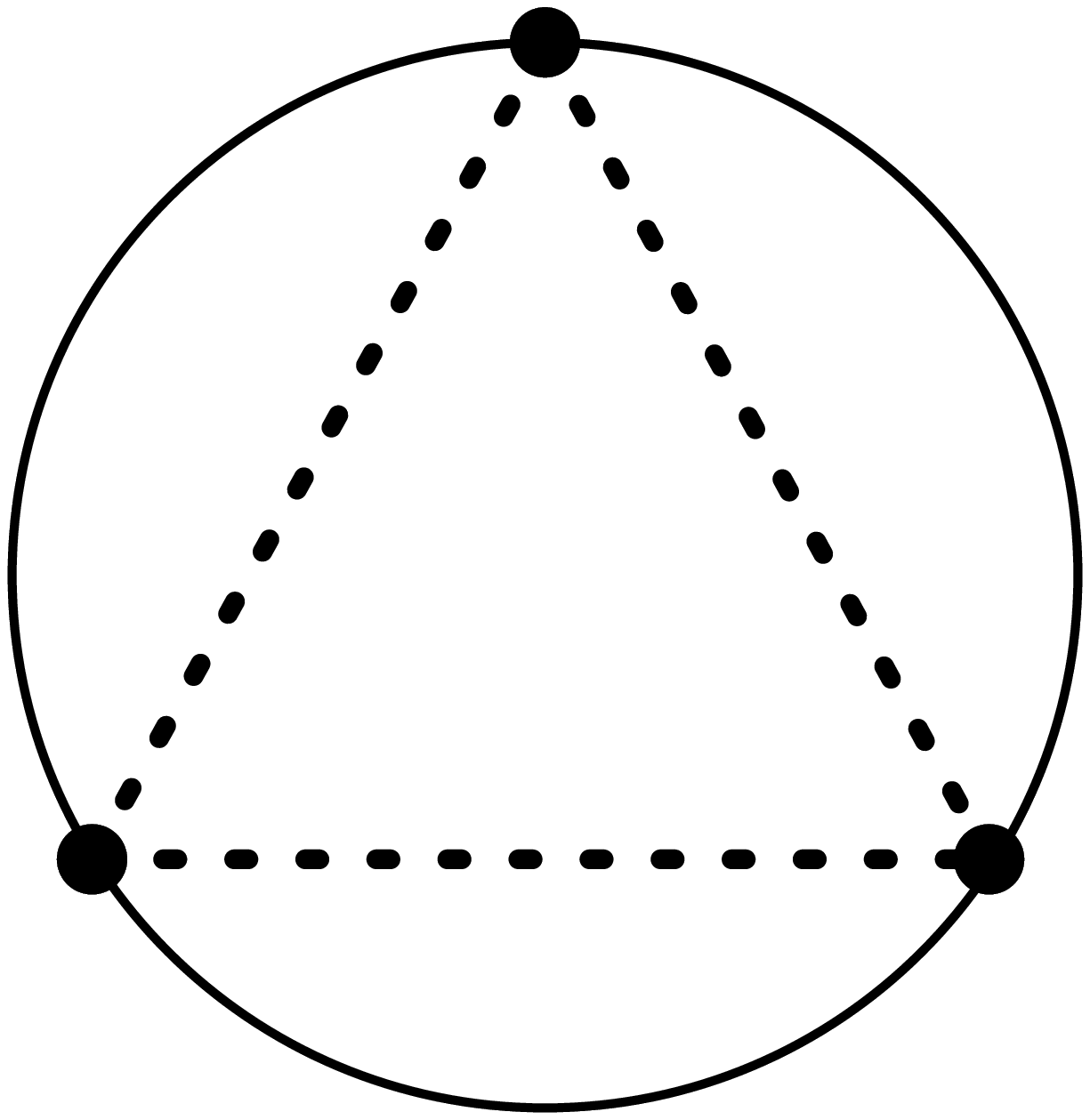}\\[0.5ex]
\hspace*{0.4cm}$\mathrm{T}_{64}$
  \end{center}
\end{minipage}
%
%
\begin{minipage}[b]{2.5cm}
  \begin{center}
    \includegraphics[height=2cm,bb=126 332 460 678]{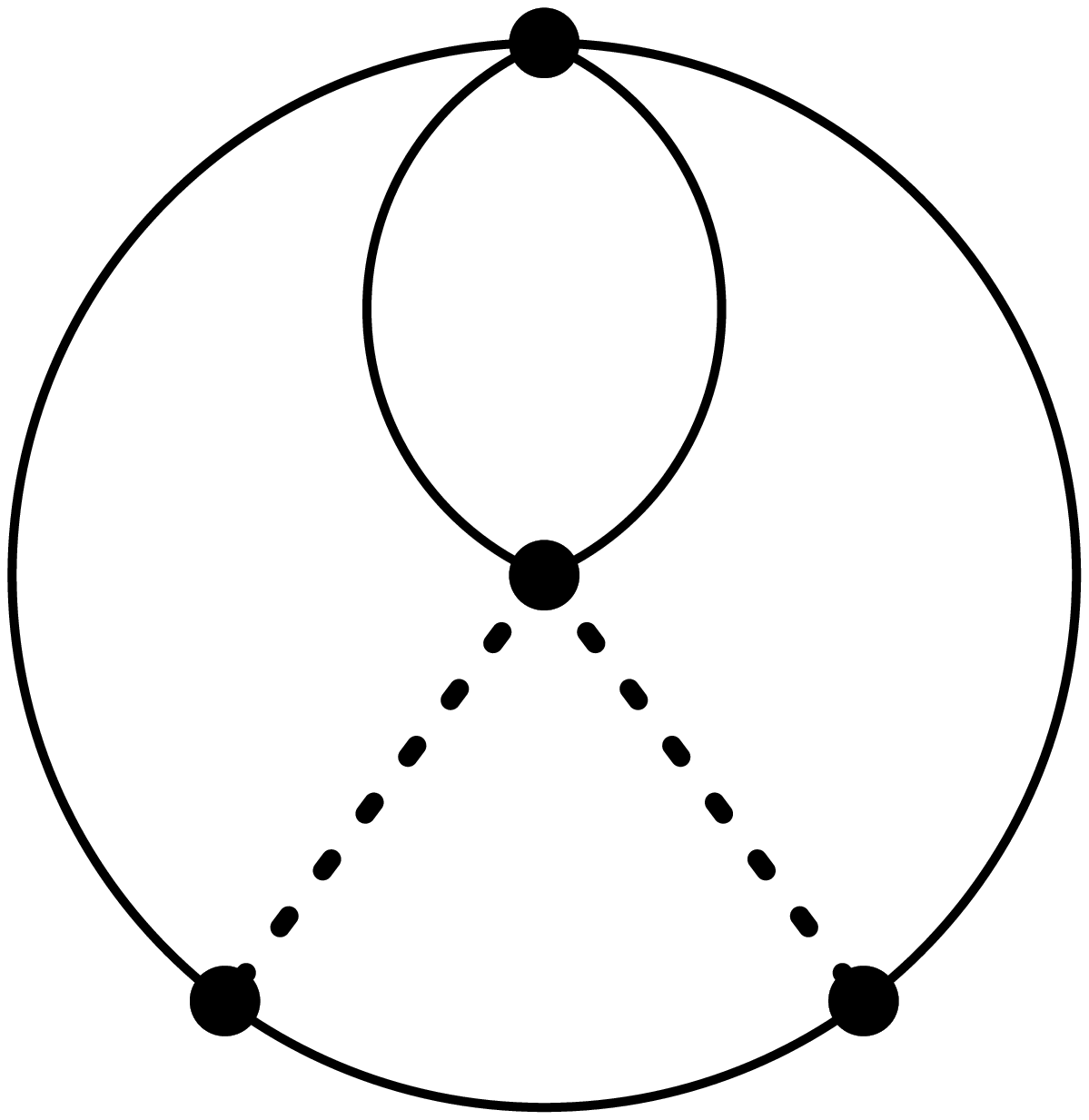}\\[0.5ex]
\hspace*{0.4cm}$\mathrm{T}_{71}$
  \end{center}
\end{minipage}
%
%
\begin{minipage}[b]{2.5cm}
  \begin{center}
    \includegraphics[height=2cm,bb=126 332 460 666]{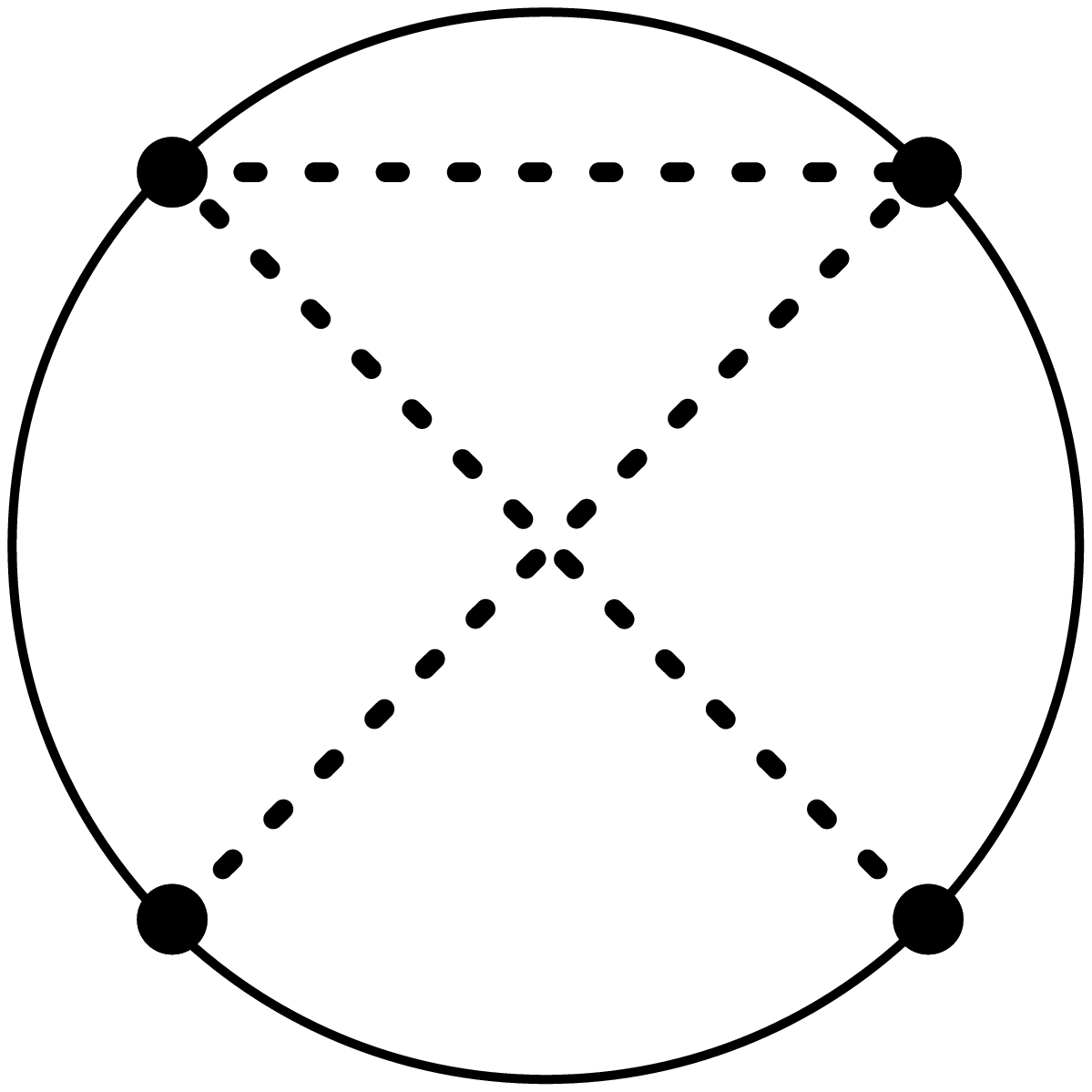}\\[0.5ex]
\hspace*{0.4cm}$\mathrm{T}_{72}$
  \end{center}
\end{minipage}
%
\\[2ex]
%
\begin{minipage}[b]{2.5cm}
  \begin{center}
    \includegraphics[height=2cm,bb=126 332 460 666]{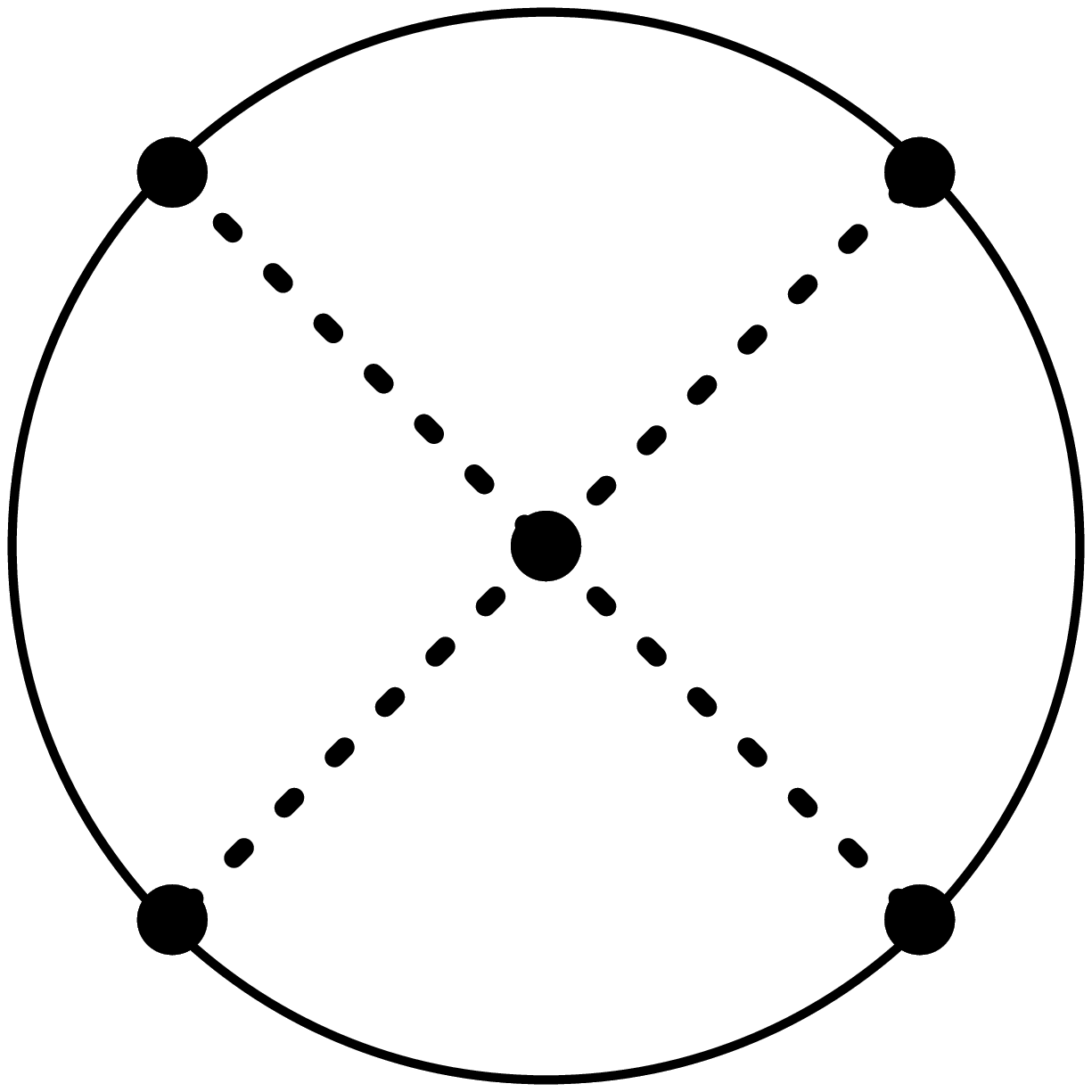}\\[0.5ex]
\hspace*{0.4cm}$\mathrm{T}_{81}$
  \end{center}
\end{minipage}
%
%
\begin{minipage}[b]{2.5cm}
  \begin{center}
    \includegraphics[height=2cm,bb=126 320 460 678]{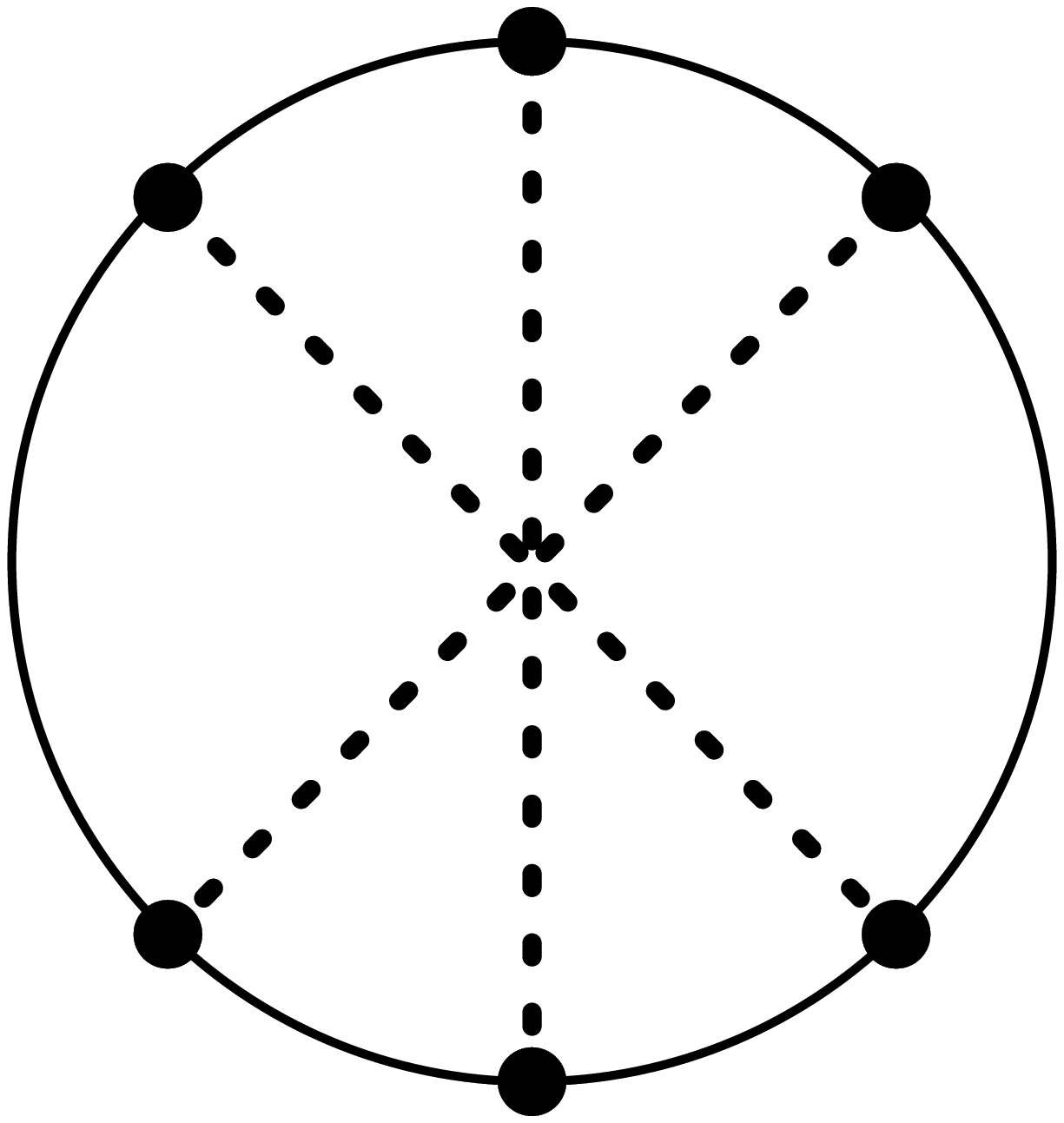}\\[0.5ex]
\hspace*{0.4cm}$\mathrm{T}_{91}$
  \end{center}
\end{minipage}
\end{center}

\caption{Master integrals where only a few terms of their
  $\vep$-expansion are known analytically. The solid (dashed) lines
  denote massive (massless) propagators.
\label{Fig:2}} 
\end{figure}

\noindent
The remaining eight master integrals (fig.~\ref{Fig:2}) have been
determined with high precision numerics with the difference equation
method \cite{Laporta:2001dd} in
ref.~\cite{Schroder:2005va}. Independently they have been determined in
ref.~\cite{Chetyrkin:2006dh} by constructing an $\vep$-finite
basis. Some of these master integrals have also been calculated in
ref.~\cite{Laporta:2002pg,Chetyrkin:2004fq,Kniehl:2005yc,Schroder:2005db}.\\
%

Inserting the master integrals and performing renormalization of the strong
coupling constant $\als$, the external current and the mass
$m=\overline{m}(\mu)$ leads to the following result ($\mu=\overline{m}$):
\begin{eqnarray}
\label{C0}
\overline{C}_0^{(3)} &=&
\nl\*\nh\*\Bigg(
 - {2\over9}\*a_4 
 + {7043\over34992} 
 - {1\over108}\*\Log{2}{4} 
\nonumber\\&&\qquad\quad
 + {\pi^2\over108}\*\Log{2}{2} 
 + {49\over12960}\*\pi^4 
 - {127\over324}\*\z3
            \Bigg)
\nonumber\\
&+& \nh^2\*\left(
   {610843\over2449440} 
 - {661\over2835}\*\z3
          \right)
\nonumber\\
&+&\nl^2\*\left(
   {17897\over69984} 
 - {31\over162}\*\z3
           \right)
\nonumber\\
&+&\nl\*\Bigg(
 - {50\over81}\*a_4 
 - {71629\over46656} 
 - {25\over972}\*\Log{2}{4} 
\nonumber\\&&\qquad
 + {25\over972}\*\Log{2}{2}\*\pi^2 
 + {8533\over116640}\*\pi^4 
 - {21343\over3888}\*\z3
         \Bigg)
\nonumber\\
&+&\nh\*\Bigg(
 - {28364\over405}\*a_4 
 - {83433703\over8164800} 
 - {7091\over2430}\*\Log{2}{4} 
\nonumber\\&&\qquad
 + {7091\over2430}\*\Log{2}{2}\*\pi^2
 + {14873\over18225}\*\pi^4 
\nonumber\\&&\qquad
 - {14509529\over340200}\*\z3 
 + {5\over3}\*\z5
         \Bigg)
\nonumber\\
&+& {64\over3}\*a_5 
 + {12007\over243}\*a_4 
 - {8572423579\over604661760} 
 + {12007\over5832}\*\Log{2}{4} 
\nonumber\\
&-&{8\over45}\*\Log{2}{5} 
 - {12007\over5832}\*\Log{2}{2}\*\pi^2 
 + {8\over27}\*\Log{2}{3}\*\pi^2
\nonumber\\ 
&-&{1074967\over699840}\*\pi^4 
 + {34\over135}\*\log{2}\*\pi^4 
 + {\pi^6\over486} 
 + {53452189\over349920}\*\z3 
\nonumber\\
&-&{28\over243}\*\z3^2 
 - {37651\over648}\*\z5  
 - {1\over432}\*T_{54,3} 
 + {7\over1296}\*T_{62,2}\,,
\end{eqnarray}
\begin{eqnarray}
\label{C1}
\overline{C}_1^{(3)} &=&
  \nl\*\nh\*\Bigg( 
 -{116\over243}\*a_4 
 + {262877\over787320}
 - {29\over1458}\*\Log{2}{4} 
\nonumber\\&&\qquad
 + {29\over1458}\*\Log{2}{2}\*\pi^2
 + {1421\over174960}\*\pi^4 
 - {38909\over58320}\*\z3
            \Bigg)
\nonumber\\
&+&\nh^2\*\left(
   {163868\over295245} 
 - {3287\over7290}\*\z3
         \right)
\nonumber\\
&+&\nl^2\*\left(
   {42173\over98415} 
 - {112\over405}\*\z3
         \right)
\nonumber\\
&+&\nh\*\Bigg(
 - {1394804\over8505}\*a_4 
 - {27670774337\over1414551600} 
\nonumber\\&&\qquad 
 - {348701\over51030}\*\Log{2}{4} 
 + {348701\over51030}\*\Log{2}{2}\*\pi^2 
\nonumber\\&&\qquad 
 + {1447057\over765450}\*\pi^4 
 - {95617883401\over943034400}\*\z3
 + {128\over27}\*\z5
        \Bigg)
\nonumber\\
&+&\nl\*\Bigg( 
 - {4793\over7290}\*a_4 
 - {9338899\over2099520} 
\nonumber\\&&\qquad
 - {4793\over174960}\*\Log{2}{4} 
 + {4793\over174960}\*\Log{2}{2}\*\pi^2
\nonumber\\&&\qquad
 + {372689\over839808}\*\pi^4 
 - {48350497\over1399680}\*\z3
         \Bigg)
\nonumber\\
&-&{127168\over1215}\*a_5 
 - {22152385\over61236}\*a_4
 + {237787820456749\over380936908800}
\nonumber\\
&-&{22152385\over1469664}\*\Log{2}{4} 
 + {15896\over18225}\*\Log{2}{5}
\nonumber\\
&+&{22152385\over1469664}\*\Log{2}{2}\*\pi^2 
 - {15896\over10935}\*\Log{2}{3}\*\pi^2 
\nonumber\\
&-&{29962031\over176359680}\*\pi^4 
 - {67558\over54675}\*\log{2}\*\pi^4
 + {60701\over1071630}\*\pi^6 
\nonumber\\
&+&{282830677079\over881798400}\*\z3
 - {242804\over76545}\*\z3^2 
 - {653339\over2430}\*\z5
\nonumber\\
&-&{5849\over272160}\*T_{54,3} 
 + {60701\over408240}\*T_{62,2}\,,
\end{eqnarray}
where Riemann's zeta-function $\zeta_n$ and the polylogarithm-function
$\mbox{Li}_n(1/2)$ are defined by
\begin{equation}
\zeta_n=\sum_{k=1}^{\infty}{1\over k^n}\quad\mbox{and}\quad
a_n=\mbox{Li}_n(1/2)=\sum_{k=1}^{\infty}{1\over2^{k}k^{n}}\,.
\label{zateunda}
\end{equation}
The two numerical constants that appear in eq.~(\ref{C0}) and
eq.~(\ref{C1}) were obtained in eq.~(19) and eq.~(20) of
ref.~\cite{Chetyrkin:2006dh}:
\begin{eqnarray}
T_{54,3}&=&-8445.8046390310298\dots \;\;\mbox{and}\;\;\nonumber\\
T_{62,2}&=&-4553.4004372195263\dots\,.
\end{eqnarray}
Numerically the coefficients $\overline{C}_0$ and $\overline{C}_1$ are
given by:
\begin{eqnarray}
\label{C0Numerik}
\overline{C}_0 &=& \left({\alpha_s\over\pi}\right)\phantom{^1} \* 1.4444
               \nonumber\\
               &+& \left({\alpha_s\over\pi}\right)^2\*
                       \left(1.5863 + 0.1387\*\nh + 0.3714\*\nl \right)
               \nonumber\\
               &+& \left({\alpha_s\over\pi}\right)^3\*
                ( 0.0252\*\nh\*\nl + 0.0257\*\nl^2 - 0.0309\*\nh^2 
               \nonumber\\&&\qquad\quad 
                 - 3.3426\*\nh - 1.2112\*\nl +1.4186 ) +\dots
               \nonumber\\
               &&\\
\mbox{and}&&\nonumber\\
\label{C1Numerik}
\overline{C}_1 &=&1.0667
               \nonumber\\
               &+& \left({\alpha_s\over\pi}\right)\phantom{^1} \* 2.5547
               \nonumber\\
               &+& \left({\alpha_s\over\pi}\right)^2\*
                     \left(0.2461 + 0.2637\*\nh + 0.6623\*\nl \right)
               \nonumber\\
               &+& \left({\alpha_s\over\pi}\right)^3\*
                ( 0.1658\*\nh\*\nl + 0.0961\*\nl^2 + 0.0130\*\nh^2 
               \nonumber\\&&\qquad\quad 
                 - 6.4188\*\nh - 2.9605\*\nl + 8.2846)+\dots\,.
               \nonumber\\
\end{eqnarray}

Using relation~(\ref{Mth}) for the first moment at three- and
four-loop approximations one can assess the influence of the new
four-loop order on the values of the charm- and bottom-quark mass.
We  first summarize the current status of the charm- and bottom-quark
masses as obtained in \cite{Kuhn:2001dm} from the first moment
evaluated to order $\alpha_s^2$:
\begin{eqnarray}
\overline{m}_c(3\GeV) &=& 1.027 \pm 0.002 \GeV
{},
\label{mc3:3l:KS}
\\
\overline{m}_b(10\GeV) &=& 3.665 \pm 0.005 \GeV
{}.
\label{mb10:3l:KS}
\end{eqnarray}
Note that here and below we display only the uncertainties coming from
the variation of the renormalization scale $\mu$ in the region $ \mu =
10 \pm 5 \GeV$ for the bottom-quark and $ \mu = 3 \pm 1 \GeV$ for the
charm-quark respectively. The experimental error  is larger and discussed in   
\cite{Kuhn:2001dm}.

Let us  discuss the influence of the newly computed
correction on the charm- and bottom-quark masses.
In our analysis we will closely follow
\cite{Kuhn:2001dm}. In particular, we will borrow from
that work  the value of the first ``experimental'' moment as
defined in eq.~(\ref{Mexp}).  The latest experimental information on
$R(s)$ which appeared after publication of \cite{Kuhn:2001dm} will be
taken into account in a future study.
The inclusion of the four-loop contribution to the function
$\overline{C}_1$ leads for the case of the charm-quark to the following
modification of eq.~(\ref{mc3:3l:KS}):
\begin{equation}
\overline{m}_c(3\GeV)  = 1.023 \pm  0.0005 \GeV
\label{mc3:4l}
{}
\end{equation}

For the case of the bottom-quark our result reads:
\begin{equation}
\overline{m}_b(10\GeV) =  3.665 \pm 0.001 \GeV
\label{mb10:4l}
{}.
\end{equation}
Thus, the four-loop correction does not change the value  of $\overline{m}_b(10\GeV)$
at all (within our accuracy) but does lead to significant decrease of the theoretical
uncertainty.

\section{Summary and Conclusion}
\label{DiscussConclude}

The calculation of Taylor expansion coefficients of the vacuum
polarization function is important for a precise determination of the
charm- and bottom-quark mass. Within this work we have presented a new
result at four-loop order in perturbative QCD for the first two
coefficients of the Taylor expansion. For the computation the
traditional IBP-method in combination with Laporta's algorithm has been
used in order to reduce all appearing integrals on a small set of master
integrals.

With the knowledge of the four-loop contributions the
theoretical uncertainty is well under control in the view of the current
and foreseeable precision of experimental data.\\

\noindent
{\bf Acknowledgments:}\\ 
The authors would like to thank M.~Faisst for numerous discussions and
for providing numerical cross checks of partial results. 
We thank M.~Czakon  for  interesting discussions. 
The work was supported by the Deutsche Forschungsgemeinschaft through the SFB/TR-9
``Computational Particle Physics''. The work of C.S. was also partially
supported by MIUR under contract 2001023713$\_$006.
\vspace{1cm}

\noindent
{\bf Note added.}
\\
\noindent
The results of our calculations as expressed in eq. (\ref{C1})
have been confirmed in the work \cite{Boughezal:2006px}, where also a
strong reduction of the theoretical error due to the unphysical scale
dependence has been found.


%
\end{document}